\documentclass[11pt]{article}
\usepackage{authblk}
\usepackage{natbib}
\usepackage{algpseudocode,algorithm}
\usepackage{listings}
\usepackage{url}
\usepackage{epsfig}
\usepackage{graphics}
\usepackage{verbatim}
\usepackage{textcomp}
\usepackage{float}
\usepackage{amssymb}
\usepackage{amsmath}
\usepackage{psfrag}
\usepackage{amscd}
\usepackage{amsfonts}
\usepackage{amsgen}
\usepackage{amstext}
\usepackage{amsthm}
\usepackage{graphicx}
\usepackage{indentfirst}
\usepackage{latexsym}
\usepackage{makeidx}
\usepackage{epsfig}
\usepackage{wrapfig}
\usepackage[all,knot,arc,import,poly]{xy}
\usepackage[usenames]{color}
\usepackage[mathscr]{eucal}
\usepackage[all]{xy}
\usepackage{verbatim}
\usepackage{rotating}
\usepackage{dsfont}
\usepackage{multirow}
\usepackage{subfigure}
\usepackage{lscape}
\usepackage[normalem]{ulem}
\usepackage{chngcntr}
\counterwithin{table}{section}
\counterwithin{figure}{section}
\usepackage{hyperref}
\usepackage[all]{hypcap}

\newtheorem{Teosub}{Theorem}[subsection]

\newtheorem{Propsub}{Proposition}[subsection]
\newtheorem{Propsubsub}{Proposition}[subsubsection]
\newtheorem{Defin}{Definition}[section]

\newtheorem{Definsubsub}{Definition}[subsubsection]
\newtheorem{Obs}{Remark}[section]
\newtheorem{Obssub}{Remark}[subsection]
\newtheorem{Obssubsub}{Remark}[subsubsection]

\newcommand*{\email}[1]{%
    \normalsize\href{mailto:#1}{#1}\par
    }

\usepackage{lipsum}
\begin{document}

\title{ \bf {Inference for Stochastically Contaminated Variable Length Markov Chains}}
\author{Denise Duarte \\ 
		\small{Departmento de Estat{\'i}stica, Universidade Federal de Minas Gerais$-$UFMG} \\
		\small{Av Antonio Carlos, 6627, Pampulha, Belo Horizonte$-$MG, Brazil}\\
    \small{\email{denise@est.ufmg.br}}							
\and Sokol Ndreca \\
		\small{Departmento de Estat{\'i}stica, Universidade Federal de Minas Gerais$-$UFMG}\\
    \small{Av. Ant\^onio Carlos 6627, Pampulha, Belo Horizonte$-$MG, Brazil}\\
    \small{\email{sokol@est.ufmg.br}}
\and Wecsley O. Prates \\ 
		\small{Departmento de Estat{\'i}stica, Universidade Federal da Bahia$-$UFBA} \\
		\small{Av Ademar de Barros  s/n, Campus de Ondina, Salvador$-$BA, Brazil}\\
		\small{\email{woprates@ufba.br}}
		}
		
		\date{}
				
	\maketitle
		
	\begin{abstract}
		In this paper, we present a methodology to estimate the parameters of stochastically contaminated models under two contamination regimes.  In both regimes, we assume that the original process is a variable length Markov chain that is contaminated by a random noise. In the first regime we consider that the random noise is added to the original source and in the second regime, the random noise is multiplied by the original source. Given a contaminated sample of these models, the original process is hidden. Then we propose a two steps estimator for the parameters of these models, that is, the probability transitions and the noise parameter, and prove its consistency. The first step is an adaptation of the Baum-Welch algorithm for Hidden Markov Models.  This step provides an estimate of a complete order $k$ Markov chain, where $k$ is bigger than the order of the variable length Markov chain if it has finite order and is a constant depending on the sample size if the hidden process has infinite order.   In the second estimation step, we propose a   bootstrap Bayesian Information Criterion, given a sample of the  Markov chain estimated in the first step, to obtain the variable length time dependence structure associated with the hidden process. We present a simulation study showing that our methodology is able to accurately recover the parameters of the models for a reasonable interval of random noises. 

\vskip.5cm

\newline 
\end{abstract}

\section{Introduction}

 In this paper, we present a new methodology to estimate the parameters of some stochastically contaminated processes. We assume that a hidden, original,  process is contaminated by some noise and only the contaminated process is observable.  In the case where the original hidden process is a Markov chain, this model is known in the literature as Hidden Markov Model (HMM) introduced in 1966 by \cite{BP}. This model has a large amount of work devoted to it due its importance and applications in subjects such as bioinformatics, communications engineering,  finance and many others. A comprehensive treatment of inference for hidden Markov models can be found in \cite{CET}. 

We analyze this problem considering that the hidden original source belongs to a larger class of process where the order of dependence in the past is not fixed as in a Markov chain.  These models are known in the literature as Variable Length Hidden Markov Models (VLHMM). As far as we know they appeared in the first time in a paper about the analysis of human movement \cite{YLJL}, \cite{YW}. In \cite{YW}, the author analyzes the 3D motion by rotating 19 major joints of the human body. The authors claim that VLHMM is superior to other models studied in relation to their efficiency and accuracy in modeling multivariate time series.

There are some previous works that analyze these class of model from a theoretical point of view, which we consider as a starting point. In \cite{GCL} the authors assume that the original source is a chain with an infinite order with a binary alphabet which is contaminated by adding to each symbol a  random Bernoulli noise, independent of the original source. In \cite{GM} the authors also consider a model where each symbol is multiplied by a Bernoulli random noise. In both articles, the authors showed that the difference between the transition probabilities of the contaminated process and the original process is limited by a constant $c$, where $c$ is a linear nondecreasing function of the random Bernoulli noise  (more details in \cite{GCL, GM}). Henceforth if the random Bernoulli noise is small enough then the contaminated sample can be used to estimate the transition probability matrix of the original hidden process.  However, if the random noise is not small enough, the approximation of the hidden transition probabilities by the estimated transition probabilities of the contaminated process is not satisfactory. Then it is crucial to estimate this noise parameter in order to know if this approximation can be applied or not. But they do not address the problem of model parameters estimation. This estimation is the main goal of the present work.

In \cite{TD} it is presented an important result in parameter estimation for a class of contaminated models similar to those analyzed in this paper. The class of models discussed in \cite{TD} is, on one hand larger than the one discussed here, since it allows random noise with a greater variety of distributions, but it is more restrictive on the other hand since only the last symbol seen in the past is considered in the conditional distributions. The author proposes an estimator based on a penalized likelihood function but, according to the author, the penalty proposed in his paper is worse than the Bayesian Information Criterion penalty, which is used in our methodology.

In this paper, we present consistent estimators for hidden parameters of the contaminated models. The simplicity of the models considered here allows us to propose an inferential methodology based on an EM algorithm for Hidden Markov Models and in the Bayesian Information Criterion (BIC). 

Besides, we present a sensitivity study of the estimators when we let the random noise to increase. Our goal with this study is to know for which interval of contamination noise the estimation procedure provides accurate estimates. 

This paper is organized as follows: Section 2 presents the basic notations and some preliminary definitions. Section 3 presents the models discussed in this paper and some theoretical results. Section 4 presents an inferential methodology and the main results for model parameter estimation. In Section 5 we perform a sensitivity simulation study concerning the influence of the random noise in the estimation procedure. Section 6 presents some conclusions. Finally, in  Appendix, we present the proofs of the results presented in Sections  3 and 4.

\section{Basic Notation and Definitions}\label{Def}
 
Let us consider a finite discrete alphabet $E=\left\{0,1,...,N-1\right\}$, with cardinality $|E|=N$. Given two integers $m,n\in\mathbb{Z}$, with $m\leq n$, we shall use the short notation $\omega_{m}^{n}$ to denote the string $(\omega_{m},...,\omega_{n})$ of symbols in $E$, and let $E^{l(\omega_{m}^{n})}$ denote the set containing such strings, where $l(\omega_{m}^{n})=n-m+1$ is the length of the string  $\omega_{m}^{n}$. An empty string is denoted by $\emptyset$ and $l(\emptyset)=0$. 

Given two strings  $\omega$ and $\upsilon$, such that $l(\omega)<\infty$, we denote by $\upsilon\omega$ the string with length $l(\upsilon)+l(\omega)$ obtained by concatenation of this two strings. The concatenation can be extended to the case when the strings are semi-infinite  $\upsilon=...\omega_{-2}\omega_{-1}$. 

We shall say that a string $\nu$  is a suffix  of the string $\omega$ if there exists a substring  $\eta$ such that $\omega=\eta\nu$. If   $1\leq l(\eta) <  l(\omega)$, $\nu$ is a proper suffix of  $\omega$ and we write
$\nu\prec\omega$. When $\nu=\omega$ we denote $\nu\preceq\omega$.

\begin{figure}
$$\xymatrix{
         &                      &                          &                         &  \bullet\ar[ld]\ar[d]  &           &\\
         &                      &                          &    \bullet\ar[ld]\ar[d] &            *+[F]{1}    &           &\\
         &                      &  \bullet\ar[ld]\ar[d]    &              *+[F]{10}  &                        &           &\\
         &     *+[F]{000}       &              *+[F]{100}  &                         &                        &           &
}$$
\caption{Context tree $\cal{T}$ with  $k=3$.}
\label{Fig2}
\end{figure}
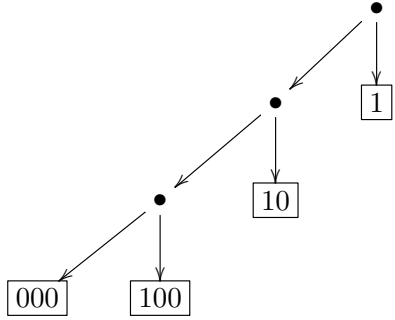

In this work we consider $\boldsymbol{X}=\left\{X_{t}\right\}_{t\in\mathbb{Z}}$ as an ergodic stochastic process on the discrete alphabet  $E$. Given an infinite string  $\omega_{-\infty}^{-1}$ and   $a\in E$, we denote by
$$p(a|\omega):=\mathbb{P}(X_{0}=a|X_{-1}=\omega_{-1},X_{-2}=\omega_{-2},...)$$
the transition  probabilities of the process $\boldsymbol{X}$ and for a finite string  $\omega\in E^{j}$, we denote by
$$p(\omega):=\mathbb{P}(X_{-j}^{-1}=\omega)$$
the initial probability distribution.
\begin{Defin}\label{Def1}
A finite string $\omega\in \cup_{j=1}^{\infty}E^{j}$  is a  {\it context} for  $\boldsymbol{X}$ if it satisfies:

\textbf{(i)} For every semi-infinite string  $x_{-\infty}^{-1}$ with $\omega$ as a suffix,
\begin{equation}\label{short}
\mathbb{P}\left(X_0=a|X_{-\infty}^{-1}=x_{-\infty}^{-1}\right)=p(a|\omega),
\end{equation}
 for every $a\in E$.

\textbf{(ii)} No proper suffix of $\omega$ satisfies (\ref{short}).
\end{Defin}
\noindent An infinite  context is a semi-infinite string  $\omega_{-\infty}^{-1}$ such that no suffix $\omega_{-j}^{-1}, \; j \in \mathbb{N}$,  is a context.
\begin{Defin}\label{Deftal}
A set $\cal{T}$ of contexts is called {\it Context Tree}, associated to the process $\boldsymbol{X}$, if no  $\omega_{1}\in\cal{T}$  is a proper suffix any other  $\omega_{2}\in\cal{T}$. A context tree satisfying   condition  \textbf{(ii):} is called {\it irreducible}.
\end{Defin}
Each context  $\omega\in\cal{T}$ can be viewed as a path from a leave to a root (see Figure \ref{Fig2}). The branches of the tree  $\cal{T}$ are identified  with the context (finite or infinite) $\omega\in\cal{T}$ in the past, the root represents the present time and it is represented by  the empty context $\emptyset$. 

Figure \ref{Fig2} shows  an order 3 context tree  $\cal{T}$ taking values  in  $E=\left\{0,1\right\}$.

\begin{Defin}\label{Defcomp}
A tree  $\cal{T}$ is complete if each  node has $|E|$ branches.
\end{Defin}
\begin{Defin}\label{Defcheia}
 If  $\cal{T}$ is  a tree such that $l(\omega)=L,\forall\omega\in\cal{T}$  then it is a order $L$ Markov chain denoted by $L_{full}$ .
\end{Defin}
\noindent We denote   $d(\cal{T}):=\max$$\left\{l \left(\omega\right):\omega\in\cal{T}\right\}$ as the depth, or order, of the tree.

A stationary stochastic process  $\boldsymbol{X}$ in $E$ is a Variable Length Markov Chain (VLMC) compatible with  a pair  $(\cal{T}$, $p(a|\omega))$ if it satisfies Definition \ref{Def1}. 
\begin{Defin}\label{Def2} Given a positive integer $k$, define the truncated tree $\cal{T}\big{|_{k}}$ of order $k$ as
\begin{center}
$\cal{T}\big{|_{k}}:=\left\{\omega\in\cal{T}\right.$: $\left. l(\omega)\leq k\right\} \bigcup \left\{\omega:l(\omega)=k\right.$ and $\omega\prec \upsilon$, for some $\left.\upsilon\in\cal{T}\right\}$.
\end{center}
\end{Defin}

\begin{Defin}\label{Defin5} A Variable Length Hidden Markov Model (VLHMM) is a bivariate stochastic process $(\boldsymbol{X},\boldsymbol{Z})$ characterized by:

\textbf{(i)}  $\boldsymbol{X}$, the hidden VLMC, with tree $\cal{T}$, taking values in the alphabet $E$;

\textbf{(ii)} $\boldsymbol{Z}$, the observable process  assuming values in  a set $O$;

\textbf{(iii)} $\bf{A}$, the transition probability matrix of the hidden process $\boldsymbol{X}$ given by $p(a|\omega)$, $\forall a \in E, \forall \omega\in\cal{T}$;

\textbf{(iv)} $\bf{B}$ (emission distribution), the conditional probability distribution for a symbol in the observable process, given the context $\omega$ in the hidden process defined by $\mathbb{P}(Z_{t}=z|X_{(t-l(\omega))+1}^{t}=\omega)$, $\forall \omega\in\cal{T}$, $\forall z \in O$;

\textbf{(v)} $\boldsymbol{\pi}$, the initial distribution of the hidden process defined by $\mathbb{P}(X^{l(\omega)}_{1}=\omega), \ \forall\omega\in\cal{T}$.
\end{Defin}
\begin{Obs}\label{Obs1}
If the hidden process  $\boldsymbol{X}$ is markovian and  $\mathbb{P}(Z_{t}=k|X_{(t-l(\omega))+1}^{t}=\omega)=\mathbb{P}(Z_{t}=k|X_{t}=j)$, that is, if the emission distribution loses memory of the whole context  then this process is an HMM,  a particular case of a VLHMM.
\end{Obs}

\section{Stochastic Contaminated Models}
 
Let us consider $\boldsymbol{X}$ a VLMC as in definition \ref{Def1} taking values in $E=\left\{0,1,\ldots,N-1\right\}, N\in \mathbb{N}$ and let  $\boldsymbol{\xi}=\left\{\xi_t\right\}_{t\in\mathbb{Z}}$ be  a sequence of independent random variables with ${\mathbb P}(\xi_t=i)=\epsilon_i$ and such that $\displaystyle\sum_{i=0}^{N-1}\epsilon_{i}=1$, independently of $\boldsymbol{X}$.  Closely following  the models presented  in \cite{GCL} and \cite{GM}, we propose two stochastic  contaminated models as follows.

\subsection{Type Sum Contamination Model}

A Type Sum Contaminated Model (TSCM) is a bivariate process  $(\boldsymbol{Z},\boldsymbol{X})$  where  $\boldsymbol{X}$ is a  VLMC, with associated tree  $\cal{T}$ and  the contaminated process $\boldsymbol{Z}$  is defined by
\begin{equation}\label{soma}
Z_{t}=X_{t}\oplus \xi_{t},
\end{equation}
where for $a, b \in E $ we define $a \oplus b= a+b\; mod(|E|)$.
\noindent The vector parameter of this model is  $\boldsymbol{\lambda}_{\boldsymbol{S}}=(\mathbf{A}_{\boldsymbol{S}},\mathbf{B}_{\boldsymbol{S}},\boldsymbol{\pi}_{\boldsymbol{S}})$, where
\begin{equation*}
\mathbf{A}_{\boldsymbol{S}}=\left\{p(a|\omega)\right\}=\mathbb{P}\left(X_0=a\bigg|X_{-l(\omega)}^{-1}=\omega\right), \forall \ a\in E, \forall \ \omega\in\cal{T}
\end{equation*}
\noindent is the transition probability matrix of the hidden process  $\boldsymbol{X}$, 
\begin{equation*}
\mathbf{B}_{\boldsymbol{S}}=\left\{b_{\omega}(z)\right\}=\mathbb{P}\left(Z_{t}=z\bigg|X_{t-l(\omega)+1}^{t}=\omega\right), \forall \ \omega\in\ E^{l(\omega)}, \forall z\in E
\end{equation*}
\noindent is the probability distribution of the observed symbol given the hidden string of the original process (emission distribution),
\begin{equation*}
\boldsymbol{\pi}_{\boldsymbol{S}}=\left\{\pi_{\omega}\right\}=\mathbb{P}\left(X_{-j}^{-1}=\omega\right), \forall \ \omega\in\cal{T}
\end{equation*}
\noindent is the initial distribution   of the original process $\boldsymbol{X}$.

 Let $\boldsymbol{\lambda_S}$ be the set of parameters of the bivariate process $(\boldsymbol{X}, \boldsymbol{Z})$. Given an observable  sample, $Z= z_1^T$,  $T\in\mathbb{N}$, the likelihood function $\mathbb{L}(\boldsymbol{\lambda_S}|Z)$ is defined by 
$$
\mathbb{L}(\boldsymbol{\lambda_S}|Z)=\mathbb{P}(\boldsymbol{\lambda_S}|Z).
$$

\begin{Obssub}We observe that in the model proposed in \cite{GCL}  $\boldsymbol{X}$ is a chain with infinite order and $E=\{0,1\}$.
\end{Obssub}
\begin{Obssub} We observe that the TSCM is a VLHMM. In consequence of the model, the emission distribution depends only on the last symbol of the context instead of the whole context. This fact allows us to propose some adaptations in the  Expectation-Maximization algorithm \cite{EM}, originally for HMM and known in the literature as the Baum-Welch algorithm, to estimate  $\boldsymbol{\lambda_{S}}$, the parameters of the VLHMM.
\end{Obssub}

The following proposition shows that: item $\bf{(i)}$  the emission distribution considering a TSCM  loses memory of the past symbols of the context; Item $\bf{(ii)}$    the number of computations needed to calculate the likelihood for a sample of a TSCM is of order $|E|^T$, where $T$ is the sample size,  and hence a direct computation is not viable. 
\begin{Propsub}\label{prop1}
\noindent Let  $\boldsymbol{Z}$ be a contaminated process as in \textbf{TSCM}.

\textbf{i)} For every $z_t,a_t,b_t\in E$ and every $\omega\in\cal{T}$, where $\cal{T}$ is the context tree of the process $\boldsymbol{X}$, the Emission distribution is 
\begin{equation}\label{eq6}
\mathbb{P}\left(Z_{t}=z_t|X_{-l(\omega)+1}^{t}=\omega\right)=\mathbb{P}\left(Z_{t}=z_t|X_{t}=a_t\right)=\mathbb{P}\left(\xi_{t}=b_t\right)I_{\left\{z_t=a_t\oplus b_t\right\}},
\end{equation}

\noindent where

\begin{center}
$I_{\left\{z_t=a_t\oplus b_t\right\}} = \left\{ \begin{array}{rl}
 1 &\mbox{ if $z_t=a_t\oplus b_t$} \\
  0 &\mbox{ otherwise.}
       \end{array} \right. $
\end{center}

\textbf{ii)} Let us consider a sample $Z=z_1^T$,  $T\in\mathbb{N}$ of the contaminated process $\boldsymbol{Z}$, such that $l(\omega)\leq T,\forall\omega\in\cal{T}$, and $k=\max\left\{l(\omega):\omega\in\cal{T}\right\}$. Then  the likelihood function $\mathbb{L}(\boldsymbol{\lambda_S}|Z)$ for the contaminated process $\boldsymbol{Z}$ can be written as:
\footnotesize{
\begin{equation}\label{eqmodsoma}
\displaystyle\sum_{a_{t},b_{t}\in E}\displaystyle\prod_{t=1}^{T}\left[\mathbb{P}(\xi_t=b_{t})\right]\left[\mathbb{P}\left(X_{1}^{k}=a_{1}^{k}\right)\displaystyle\prod_{t=k+1}^{T}\mathbb{P}\left(X_{t}=a_{t}|X_{t-l(\omega)}^{t-1}=a_{t-l(\omega)}^{t-1}\right)\right]\displaystyle\prod_{t=k+1}^{T}I_{\left\{z_{t}=a_{t}\oplus b_{t}\right\}}.
\end{equation}
}\normalsize
\end{Propsub}
\bigskip

\textbf{Proof} See Appendix.

\begin{Obssub}
If the  VLMC $\boldsymbol{X}$ has infinite order, we can obtain, in an analogous way, a truncated version of the likelihood function $\mathbb{L}(\boldsymbol{\lambda_S}|Z)$ at some finite order $L< T$.
\end{Obssub}

\begin{Obssub}

Due to item $\bf{i)}$ the emission distribution $ \mathbf{B}_{\boldsymbol{S}}= \mathbf{B}_{\boldsymbol{S}} (\epsilon)$ depends only on the random noise.
\end{Obssub}

\subsection{Type Product Contaminated Model}\label{mod2}

We define a Type Product  Contaminated Model (TPCM)  also as a bivariate process ($\boldsymbol{Z},\boldsymbol{X}$) where $\boldsymbol{X}$ is a VLMC and the contaminated process $\boldsymbol{Z}$ is obtained  by
\begin{equation}\label{eq0101}
Z_{t}=X_{t}\cdot\xi_{t}.
\end{equation}
\noindent We denote the vector of parameters of this model by $\boldsymbol{\lambda}_{\boldsymbol{P}}=(\mathbf{A}_{\boldsymbol{P}},\mathbf{B}_{\boldsymbol{P}},\boldsymbol{\pi}_{\boldsymbol{P}})$, where 
\begin{equation*}
\mathbf{A}_{\boldsymbol{P}}=\left\{p(a|\omega)\right\}=\mathbb{P}\left(X_t=a\bigg|X_{-l(\omega)}^{-1}=\omega\right), \forall \ a\in E, \forall \ \omega\in\cal{T}
\end{equation*}
\noindent is the transition probability matrix of the hidden process $\boldsymbol{X}$,
\begin{equation*}
\mathbf{B}_{\boldsymbol{P}}=\left\{b_{\omega}(z)\right\}=\mathbb{P}\left(Z_{t}=z\bigg|X_{t-l(\omega)+1}^{t}=\omega\right), \forall \ \omega\in\ E^{l(\omega)}, \forall z\in\ E
\end{equation*}
\noindent is the probability distribution of the observed symbol given the hidden string of the original process (emission distribution), 
\begin{equation*}
\boldsymbol{\pi}_{\boldsymbol{P}}=\left\{\pi_{\omega}\right\}=\mathbb{P}\left(X_{-j}^{-1}=\omega\right), \forall \ \omega \in\cal{T}
\end{equation*}
is the initial distribution of the original process $\boldsymbol{X}$. 

\bigskip
\begin{Obssub} The model proposed in \cite{GM} is a  TPCM with  $E= \{0,1\}$. 
\end{Obssub}

\begin{Obssub} Observe that the TPCM is also a VLHMM and, also as consequence of the model, the emission distribution depends only on the last symbols of the context and not on the whole context. 
\end{Obssub}

In the same way as in TSCM, the following proposition shows that: Item $\bf{(i)}$, the emission distribution considering a TPCM also loses memory of the past symbols of the context; Item $\bf{(ii)}$,   the number of computations needed to calculate the likelihood for a sample of a TSCM is also of order $|E|^T$, where $T$ is the sample size,  and hence a direct computation is not viable.

\begin{Propsub}\label{propmod2} 
Let $\boldsymbol{Z}$ be a contaminated process as  in a  TPCM.\\
\textbf{i)} For every $z_t,a_t,b_t\in E$ and every $\omega\in\cal{T}$, where $\cal{T}$ is the context tree of the process $\boldsymbol{X}$, the emission distribution is 
\begin{equation}\label{eqmod26}
\mathbb{P}\left(Z_{t}=z_t|X_{-l(\omega)+1}^{t}=\omega\right)=\mathbb{P}\left(Z_{t}=z_t|X_{t}=a_t\right)=\mathbb{P}\left(\xi_{t}=b_t\right)I_{\left\{z_t=a_t.b_t\right\}}.
\end{equation}

\textbf{ii)} Let us consider a sample $Z=z_1^T$, $T\in\mathbb{N}$  of the contaminated process $\boldsymbol{Z}$,  such that $l(\omega)\leq T,\forall\omega\in\cal{T}$, and $k=\max\left\{l(\omega):\omega\in\cal{T}\right\}$. Then  the likelihood function $\mathbb{L}(\boldsymbol{\lambda_P}|Z)$ for the contaminated process $\boldsymbol{Z}$ can be written as:
\footnotesize{
\begin{equation}\label{eqmodprod}
\displaystyle\sum_{a_{t},b_{t}\in E}\displaystyle\prod_{t=1}^{T}\left[\mathbb{P}(\xi_t=b_{t})\right]\left[\mathbb{P}\left(X_{1}^{k}=a_{1}^{k}\right)\displaystyle\prod_{t=k+1}^{T}\mathbb{P}\left(X_{t}=a_{t}|X_{t-l(\omega)}^{t-1}=a_{t-l(\omega)}^{t-1}\right)\right]\displaystyle\prod_{t=1}^{T}I_{\left\{z_{t}=a_{t}. b_{t}\right\}}.
\end{equation}
}\normalsize

\end{Propsub}
\textbf{Proof} See Appendix.

\begin{Obssub}
If the  VLMC $\boldsymbol{X}$ has infinite order, we can obtain, in an analogous way, a truncated version of the  likelihood function $\mathbb{L}(\boldsymbol{\lambda_P}|Z)$ at some finite order $L< T$.
\end{Obssub}

\begin{Obssub}

Due to item $\bf{i)}$, $ \mathbf{B}_{\boldsymbol{P}}= \mathbf{B}_{\boldsymbol{P}} (\epsilon)$,  the emission distribution, depends only on the random noise.
\end{Obssub}

\section{Inference for TSCM and TPCM }

Considering that a direct computation of the likelihood is intractable, we propose an EM  algorithm, based on the Baum-Welch algorithm for HMM,  to iteratively compute the likelihood for VLHMM.  To this end, we propose a new parameterization of the models as follows.

Let us  first consider a finite VLHMM $(\boldsymbol{X},\boldsymbol{Z})$, where $\boldsymbol{X}$ has a finite tree $\cal{T}$ and let $k$ be the length of the biggest context,  $k=\max\left\{l(\omega):\omega\in\cal{T}\right\}$. We can rewrite the order $k$ VLHMM as  an order $k$ Markov chain as
$ \boldsymbol{X^*}= \left\{X^*_r\right\}_{r \in \mathbb{N}}$,   which is a  $k_{full}$ tree, assuming values in $E^k$ where
$$X^*_{r}:={X}_{(r+k)-1}^{r}, \ r=1,...,(T-k)+1, \ T\in\mathbb{N}.$$
The transitions probabilities of $\boldsymbol{X^*}$ are defined by $\boldsymbol{A^*}=\left\{p^*(\omega|\nu)\right\}, \forall \omega,\nu\in\ E^{k}$ and with initial distribution $\boldsymbol{\pi^*}=\left\{\mathbb{P}(X^*_{1}=\omega)\right\}, \forall \omega\in\ E^{k}$. 

Similarly we define a new observable process  $ \boldsymbol{Z^*}= \left\{Z^*_r\right\}_{r \in \mathbb{N}}$, assuming  values in $ E^{k}$, where
$${Z}^{*}_{r}={Z}_{(r+k)-1}^{r}, \ r=1,...,(T-k)+1.$$

In this way, the VLHMM $(\boldsymbol{X},\boldsymbol{Z})$ can be viewed as an HMM $(\boldsymbol{X^*},\boldsymbol{Z^{*}})$ with set of parameters $\boldsymbol{\lambda^{*}}=(\boldsymbol{A^*},\boldsymbol{B^*},\boldsymbol{\pi^* })$.

As an example, suppose that $\boldsymbol{X}$ is a VLMC that assumes values in the alphabet $E=\left\{0,1\right\}$ and $k=\max\left\{l(\omega):\omega\in\cal{T}\right\}=2$. Given a sample of the hidden process $\boldsymbol{X}$,  $x_1^T=\left\{0,0,1,0,1,1,0,1,...,0,1\right\}$,  then  the sample associated to Markov chain $\boldsymbol{X^*}$ with order $k=2$, denoted by $X^*$, is 
$$X^*= x_1^{r *}=\left\{00,01,10,01,11,10,01,...,01\right\}$$
And  for an observable  sample  $z_1^T=\left\{0,0,1,1,0,0,1,...,1,0\right\}$,  of the contaminated process $\boldsymbol{Z}$,  we have that the new observable sample of the process $\boldsymbol{Z^{*}}$ is  
$$Z^{*}=z_1^{r*}=\left\{00,01,11,10,00,01,...,10\right\}$$

Now we are ready to apply the Baum-Welch EM algorithm \cite{BP}  to the process $(\boldsymbol{X}^*, \boldsymbol{Z}^* )$. Given a sample $Z^{*}$, the forward variable is defined as

\begin{center}
$\alpha_{r}\displaystyle\left(\omega\right)=\mathbb{P}\left(z^{*}_{1},...,z^{*}_{r},X^*_{r}=\omega\bigg|\boldsymbol{\lambda}^{*}\right).$
\end{center}
By induction, we have that
\begin{center}
$\alpha_{1}\left(\omega\right)=\pi_{\omega}b_{\omega}\left(z^{*}_{1}\right), \forall \omega\in E^{k},$
\end{center}
\begin{center}
$\alpha_{r+1}\left(\omega\right)=\left[\displaystyle\sum_{\nu \in E^k}\alpha_{r}\left(\omega\right)p^*\left(\omega|\nu\right)\right]b_{\omega}\left(z^{*}_{r+1}\right), \forall\omega\in E^{k},\ \ 2\leq r\leq (T-k)+1$.
\end{center}
Similarly,  the backward variable is defined as
\begin{center}
$\beta_{r}\left(\omega\right)=\mathbb{P}\left(z^{*}_{r+1},z^{*}_{r+2},...,z_{(T-k)+1}|X^*_{r}=\omega,\boldsymbol{\lambda}^{*}\right),$
\end{center}
and by induction follows
\begin{center}
$\beta_{(T-k)+1}\left(\omega\right)=1, \forall \omega\in E^{k},$
\end{center}
\begin{equation*}
\beta_{r}\left(\omega\right)=\displaystyle\sum_{\nu \in E^{k}}p^*\left(\omega|\nu\right)b_{\omega}\left(z^{*}_{r+1}\right)\beta_{r+1}\left(\omega\right), \forall \omega\in E^{k},\ \ r=(T-k)+2,(T-k)+3,...,1.
\end{equation*}

\noindent Now we define 
\begin{center}
$\gamma_{r}\left(\omega\right)=\mathbb{P}\left(X^{*}_{r}=\omega|Z^{*},\boldsymbol{\lambda}^{*}\right),$
\end{center}

and 
\begin{equation*}
\delta_{r}\left(\omega,\nu\right)=\mathbb{P}\left(X^*_{r}=\omega,X^*_{r+1}=\nu|Z^{*},\boldsymbol{\lambda}^{*}\right).
\end{equation*}

Given $\alpha_{r}\displaystyle\left(\omega\right)$ and $\beta_{r}\displaystyle\left(\omega\right)$,
 we can write

\begin{equation}\label{eqgama}
\gamma_{r}\left(\omega\right)=\frac{\alpha_{r}\left(\omega\right)\beta_{r}\left(\omega\right)}{\displaystyle\sum_{\omega \in E^{k}}\alpha_{r}\left(\omega\right)\beta_{r}\left(\omega\right)},
\end{equation}

and
\begin{equation*}\label{eqdelta}
\delta_{r}\left(\omega,\nu\right)=\frac{\alpha_{r}\left(\omega\right)p^*\left(\omega|\nu\right)b_{\omega}\left(z^{*}_{r+1}\right)\beta_{r+1}\left(\omega\right)}{\displaystyle\sum_{\omega \in E^{k}}\displaystyle\sum_{\nu \in E^{k}}\alpha_{r}\left(\omega\right)p^*\left(\omega|\nu\right)b_{\omega}\left(z^{*}_{r+1}\right)\beta_{r+1}\left(\omega\right)}, \ \forall \omega,\nu\in E^{k}.
\end{equation*}

Henceforth, the parameter vector $\boldsymbol{\lambda}^*$ can be updated in the follwoing way:

\begin{center}
$\hat{\boldsymbol{\pi}}^*=\left\{\hat{\pi}_{\omega}^{*}\right\}_{\forall {\omega\in E^{k}}}=\left\{\gamma_{1}(\omega)\right\},$
\end{center}

\begin{center}
$\hat{\mathbf{A}}^*=\left\{\hat{p}^*(\omega|\nu)\right\}_{ \forall  {\omega,\nu\in E^{k}}},$ where 
$\hat{p}^*(\omega|\nu)= \frac{\displaystyle\sum^{T+k-1}_{r=1}\delta_{r}(\omega,\nu)}{\displaystyle\sum^{T+k-1}_{r=1}\gamma_{r}(\omega)}$, 
\end{center}
\begin{center}
$\hat{\mathbf{B}}^*=\left\{\hat{b}_{\omega}(\nu)\right\}_{ \forall {\omega,\nu\in E^{k}}}$, where $\hat{b}_{\omega}(\nu)=\frac{\displaystyle\sum^{T+k}_{r=1} I_{\left\{z^{*}_r=\nu\right\}}\gamma_{r}(\omega)}{\displaystyle\sum^{T+k}_{r=1}\gamma_{r}(\omega)}.$
\end{center}

\begin{Obs} If   $\boldsymbol{X}$ has infinite order, for a finite sample $z_1^T$ it is possible to estimate only a truncated tree $\cal{T}\big{|_k}$, where $k\in\mathbb{N}$ is as big as possible. We apply the same methodology proposed for finite trees to obtain $\hat{\cal{T}}\big{|_k}$ using a truncated likelihood function, given the sample of the observable process $\boldsymbol{Z}$.

\end{Obs}

\subsection{Estimation Methodology}

First of all, we stress that, for estimation purposes, the main difference between a VLHMM and an HMM is that in an HMM the states of the original (hidden) process are known while in a VLHMM they are unknown. Then the estimation procedure has also to learn the contexts of the hidden tree  $\cal{T}$ associated to the VLMC $\boldsymbol{X}$. This fact makes the estimation procedure much more complex. In order to overcome this difficulty, we  propose a methodology that has two steps described as follows. We describe the procedure considering a  finite order VLMC. The infinite order case is analogous.

\begin{enumerate} 

\item {\it First step: } Given a sample $z_1^T$ of the obervable process, for  $k$ fixed, which  depends on the sample size $T$, estimate the hidden  $k_{full}$ Markov chain associated to the order $k$ VLMC, $\hat{\cal{T}}_k$, applying the  Baum-Welch algorithm considering that  the process is the HMM  $\left( \boldsymbol{X^*},  \boldsymbol{Z^*} \right)$. 

\item {\it Second step:}  Generate a bootstrap sample from the estimated tree
$\hat{\cal{T}}_k$, with transition matrix $\hat{A}^*$, and 
apply a pruning  procedure  based on a Bayesian Information Criterion to obtain the estimated tree $\hat{\cal{T}}$ of the $\cal{T}$.

\end{enumerate}

 In the first step, since the Baum-Welch (BW) algorithm is an EM algorithm, its convergence to a local maximum of the likelihood function is guaranteed \cite{BP}. 
Our propose to make the BW  algorithm to reach a global maximum is to vary the required initial guess $\boldsymbol{\lambda}^{*}_0$  over a set

\begin{center}$\Lambda_0^*= \left\{\boldsymbol{\lambda}^{*1}_{0}, \ldots,\boldsymbol{\lambda}^{*N}_{0}\right\},$\end{center} for a fixed $N$, as big as possible. 

Then, for each initial guess, $\boldsymbol{\lambda}^{*j}_0$,  the Baum-Welch algorithm returns an estimate $\tilde{\boldsymbol{\lambda}}_j^{*}\in \Lambda= \{\tilde{\boldsymbol{\lambda}}_1^{*}, \ldots, \tilde{\boldsymbol{\lambda}}_N^{*}\} $ of the vector of parameters $\boldsymbol{\lambda}^{*}=\left(\boldsymbol{A^*},\boldsymbol{B^*}, \boldsymbol{\pi}^{*}\right)$.

Finally, our propose  to estimate  $\boldsymbol{\lambda^{*}}$ is to take the $\tilde{\boldsymbol{\lambda}}^{*}_j$   that  maximizes the likelihood $\mathbb{L}\left(\tilde{\boldsymbol{\lambda}}^{*}|Z^*\right)$, given the observed sample $Z^*=z^{*}_{1},...,z^*_{(T-k)+1}$, that is 
\begin{equation}\label{vero}
\hat{\boldsymbol{\lambda}}^{*}=\arg\displaystyle\max_{\tilde{\boldsymbol{\lambda}}^{*}\in\boldsymbol{\Lambda}}\mathbb{L}\left(\tilde{\boldsymbol{\lambda}}^{*}|Z^*\right).
\end{equation}

In this way,  if the likelihood function $ \mathbb{L}\left(\tilde{\boldsymbol{\lambda}}^{*}|Z^*\right)$ has a finite number of local maxima, for $N$ large enough, our estimator $\hat{\boldsymbol{\lambda}}^{*}$ is  the Maximum Likelihood Estimator (MLE) of ${\boldsymbol{\lambda}}^{*}$.

  In the estimation procedure, for each $\boldsymbol{\lambda}^{j*}_0$,  $j \in \{1, \ldots, N\}$, we consider that the values of the noise parameter $\epsilon _j, j \in\{1, \ldots, N\}$ in the initial emission distribution   $ \mathbf{B}_0(\epsilon_j)$ assumes a distinct increasing value in the interval  $ (0,1)$. Then,  for each value of the  noise parameter, $\epsilon_{j}$, it corresponds  an initial vector $\boldsymbol{\lambda}^{*}_0(\epsilon_j)$. Henceforth, $\hat{\boldsymbol{\lambda}}^{*}$ also  provides an estimator $\hat{\epsilon}$ of the noise parameter since we can rewrite (\ref{vero}) in terms of $\epsilon_{j}$. This estimator is defined in the following way 
  
  \begin{equation}\label{veroepsilon}
\hat{\epsilon}=\arg\displaystyle\max_{\epsilon_j }\mathbb{L}\left(\tilde{\boldsymbol{\lambda}}^{*}(\epsilon_j)|Z^*\right).
\end{equation}

We keep fixed the initial guess of the transition probabilities of the $k_{full}$ (matrix) $\boldsymbol{A^*}$, for each $j \in \{1, \ldots, N\}$,  as  the empirical transition matrix of the observable values $\boldsymbol{Z^{*}}$, truncated at  order $k$. And $\boldsymbol{\pi}^{*j}_0$  is considered as an uniform distribution.

\bigskip

 Now we proceed with the second step of our estimation procedure.  Once we have the estimate $\hat{\boldsymbol{\lambda}^{*}}=\left(\hat{\boldsymbol{A^*}},\hat{ \boldsymbol{B^*}}, \hat{\boldsymbol{\pi}^*}\right)$ of $\boldsymbol{\lambda^{*}}=\left(\boldsymbol{A^*}, \boldsymbol{B^*}, \boldsymbol{\pi}^{*}\right)$, associated to the HMM $(\boldsymbol{X^*}, \boldsymbol{Z^*})$, we propose a pruning in the estimated $k_{full} $ tree,  $\hat{\cal{T}}_{k}$, with transition matrix $ \hat{\boldsymbol{A}^*}$,  to obtain an estimator  for the parameters of the context tree $\cal{T}$, associated to the hidden process  $\boldsymbol{X}$. To this end, we apply an adaptation of the  Bayesian Information Criterion (BIC) estimator for VLMC,  proposed in \cite{CT}, which is explained in this section. Under some mild conditions, \cite{CT} showed that the BIC pruning  provides a consistent estimator for a VLMC when the sample comes from a VLMC. However, this is not the case here, since we do not have a sample directly from the hidden process. Then we propose  a pruning procedure based on  a sample  of the estimated $k_{full}$ tree $\hat{\cal{T}}_{k}$, with transition matrix $ \hat{\boldsymbol{A}^*}$. Thus, our proposal is a bootstrap version of the BIC algorithm, where we replace a sample of the true  VLMC by a bootstrap sample ${\hat{x}_1^m}:= \hat{x}_1, \ldots,\hat{ x}_m, \ m=O(T),$ drawn from the estimated transition matrix $\hat{\mathbf{A}}^*$. 

Following \cite{CT}, we need to define some auxiliary variables. Let $\hat{N}_m(\omega,a)$ be the number of occurrences of the string $\omega\in\cup_{j=1}^k E^{j}$ followed by the symbol  $a \in E$ in the bootstrap sample $\hat{x}_1^m$, that is

$$\hat{N}_m(\omega,a)=\left|\left\{i:D(m) <i\leq m,{\hat{x}}^{i-1}_{i-l(\omega)}=\omega,{\hat{x}}_i=a\right\}\right|$$

and the number of occurrences of $\omega$ in $\hat{x}_{1}^m$ is  

$$\hat{N}_m(\omega)=\left|\left\{i:D(m)<i\leq m,{\hat{x}}^{i-1}_{i-l(\omega)}=\omega\right\}\right|.$$

 A feasible bootstrap context tree is such that, given a bootstrap   sample  $\hat{x}_1^m$, $d({\cal{T}})\leq$ $D(m)$,  $\hat{N}_m(\omega)\geq 1$ for all $\omega\in {\cal{T}}$, and  $\omega^{\prime}$ is a  suffix of some  $\omega \in \cal{{T}}$ with  $\hat{N}_m(\omega)^{\prime}\geq 1$. The set of boostrap feasible context trees is denoted by
$ \mathcal{F}\mbox{\scriptsize$(\hat{x}_1^{m},D(m))$}.$

We define the {\it Bootstrap Bayesian Information Criterion} (BIC) for a set of feasible trees  as 

\begin{equation}\label{bictrue}
BIC_{\cal{T}}(\hat{x}_1^m)= -\log ML_{\cal{T}}(\hat{x}_1^m) + \frac{(|E|-1)|\cal{T}|}{2} \log m,
\end{equation}
where $ML_{\cal{T}}(\hat{x}_1^m)=\displaystyle\prod_{\omega\in\tau: \hat{N}_m(\omega)\geq 1}
\displaystyle\prod_{a\in E}\left(\frac{\hat{N}_m(\omega,a)}{\hat{N}_m(\omega)}\right)^{\hat{N}_m(\omega,a)}$.

\bigskip

For   a finite  bootstrap sample, $\hat{x}_1^m$,   the  BIC estimator of $\cal{T}$ is defined by 
\begin{equation}\label{bic}
\hat{\cal{T}}_{BIC}\left(\hat{x}_{1}^{m}\right)=\arg\displaystyle\min_{\cal{T}\in \mathcal{F}\mbox{\scriptsize$(\hat{x}_1^{m},D(m))$}}BIC_{\cal{T}}(\hat{x}_1^{m}), \; \;,
\end{equation}
with $D(m)=o(\log m)$.

 Since we have replaced the sample of the VLMC by a bootstrap sample we need to show that the  BIC estimator is still consistent. To this end, we present the following proposition which plays a key role to prove this consistency.

\begin{Propsub}\label{coroCT} Let $\hat{\mathbf{A}}^*$ be a strongly consistent estimator of the transition probability matrix of the markovian process $\boldsymbol{X}^*$, with law $\hat{P}$ and transition probabilities $\hat{p}(a,\omega)$, $\forall a \in E, \omega \in \hat{\cal{T}}_k$. And let $\hat{x}_1^m$ be a bootstrap sample of size $m=O(T)$, drawn from $\hat{P}$ fixed, where $T$ is the size of the hidden sample, $x_1^T$. Then, conditionally on $\hat{P}$,  $\forall a \in E, \omega \in \hat{\cal{T}}_k$ and  for almost all realizations of the VLHMM $\left(\boldsymbol{X}, \boldsymbol{Z}\right)$,

$$
\frac{\hat{N}_m(\omega a)}{\hat{N}_m(\omega)} \longrightarrow p(a|\omega) , \; almost\; surely \;as\; T \rightarrow\infty.$$
\end{Propsub}

\noindent{\bf Proof in Appendix.}

\bigskip

\noindent Now we are ready to state the main result of this work.
\begin{Teosub}\label{teoCT}
Let  $\hat{x}_1^m$ be a bootstrap sample of size  $m=O(T)$ drawn from $\hat{P}$ fixed. For $d(\cal{T})< \infty$,  for the BIC   estimator of  $\cal{T}$, given by equation (\ref{bictrue}) we have that

\begin{center}$\hat{\cal{T}}_{BIC}\left(\hat{x}_1^m\right)=\cal{T},$\end{center}
almost surely when  $m \rightarrow \infty$.

\noindent In the general case, we have
\begin{center}$\hat{\cal{T}}_{BIC}\big{|_k}\left(\hat{x}_1^m\right)=\cal{T}\big{|_k},$\end{center}
almost surely when  $m \rightarrow \infty.$
\end{Teosub}

\noindent{\bf Proof in Appendix.}

\subsubsection{Computation of the Bootstrap BIC Estimator}

The direct application of the BIC procedure is impracticable due to a large number of possible trees to be checked in the likelihood function. Then the estimation of the tree $\cal{T}$, associated to the hidden process $\boldsymbol{X}$, is made in the same manner proposed in \cite{CT} through a recursive procedure, based on the CTM algorithm \cite{CT}, which assigns a value and a binary indicator to each node.  The difference in our procedure is that the sample is not directly generated from a VLMC but from an estimated tree $\hat{\cal{T}}_k$. In the following we adapt the definitions given in \cite{CT}, replacing the original sample by the bootstrap sample $\hat{x}_1^m$. Let

\begin{equation*}
\tilde{\mathbb{P}}_{L,\omega}(\hat{x}_1^m)=\left\{\begin{array}{lcr}\displaystyle\prod_{a \in E}\left(\frac{\hat{N}_m(\omega,a)}{\hat{N}_m(\omega)}\right)^{\hat{N}_m(\omega,a)}& \mbox{if} & \hat{N}_m(\omega)\geq 1, \\ 1 & \mbox{if} & \hat{N}_m(\omega)=0. \end{array}\right.
\end{equation*}
\noindent Then  the estimator $\hat{\cal{T}}_{BIC}\left(\hat{x}_1^m\right)$, defined in equation (11), can be written as
\begin{equation}\label{BIC}
\hat{\cal{T}}_{BIC}\left(\hat{x}_1^m\right)=\arg\displaystyle\max_{\cal{\hat{T}}\in \mathcal{F_{\mbox{\tiny 1}}}\mbox{\scriptsize$(\hat{x}_1^m,D(m))$}}\displaystyle\prod_{\omega \in \cal{\hat{T}}}\tilde{\mathbb{P}}_\omega\left(\hat{x}_1^m\right),
\end{equation}
\noindent where $\tilde{\mathbb{P}}_\omega\left(\hat{x}_1^m\right)=m^{-\frac{\left|E\right|-1}{2}}\tilde{\mathbb{P}}_{L,\omega}(\hat{x}_1^m)$.

\begin{Definsubsub} Given a sample $\hat{x}_1^m$, let $S_d$ be the set of all contexts of maximum size $d=D(m)=o(log m)$ and  such that  $\hat{N}_m(\omega)\geq 1$. For each string $\omega\in S_d$ with $\hat{N}_m(\omega)\geq 1$, we assign recursively starting from the leaves of the $d_{full}$ tree $\hat{\cal{T}}_d$, the value
\begin{equation*}\label{vari1}
V_{\omega}^{d}(\hat{x}_1^m)=\left\{\begin{array}{lcr}\max\left\{\tilde{\mathbb{P}}_\omega(\hat{x}_1^m),\displaystyle\prod_{a \in E:N_T^{\hat{x}_1^m}(a\omega)\geq 1}V^d_{a\omega}(\hat{x}_1^m)\right\} & \mbox{if} & 0\leq l(\omega)<d \\ \tilde{\mathbb{P}}_\omega(\hat{x}^m_1) & \mbox{if} & l(\omega)=d.\end{array}\right.
\end{equation*}
and the indicator function
\begin{equation*}\label{vari2}
\mathcal{X}^{d}_{\omega}(\hat{x}_1^m)=\left\{\begin{array}{lcr}1 & \mbox{if} & 0\leq l(\omega)<d, \displaystyle\prod_{a \in E:\hat{N}_m(a\omega)\geq 1}V^d_{a\omega}(\hat{x}_1^m)>\tilde{\mathbb{P}}_\omega(\hat{x}_1^m) \\ 0 & \mbox{if} & 0 \leq l(\omega)<d, \displaystyle\prod_{a \in E:\hat{N}_m(a\omega)\geq 1}V^d_{a\omega}(\hat{x}_1^m)\leq\tilde{\mathbb{P}}_\omega(\hat{x}_1^m) \\ 0 & \mbox{if} & l(\omega)=d.\end{array}\right.
\end{equation*}
\end{Definsubsub}

\begin{Definsubsub}

For each  $\omega \in S_d$ the estimated tree $\hat{\cal{T}}$ is the set of contexts $\nu\succeq \omega$ such that\\

$$
\hat{\cal{T}}^{d}_{\omega}\left(\hat{x}_1^m\right) := \left\{\begin{array}{rl}
\{\nu \in S_d:\mathcal{X}^{d}_{\nu}(\hat{x}_1^m)=0, \ \mathcal{X}^{d}_{\upsilon}(\hat{x}_1^m)=1, \ \forall \omega\preceq \upsilon\preceq \nu\}, &\mbox{ if }\mathcal{X}^{d}_{\omega}(\hat{x}_1^m)=1 \\ \noindent
 \left\{\omega\right\}, &\mbox{ if }\mathcal{X}^{d}_{\omega}(\hat{x}_1^m)=0
       \end{array} \right.
$$
\end{Definsubsub}

\begin{Propsubsub}\label{root}
The bootstrap context tree estimator $\hat{\cal{T}}_{BIC}\left( \hat{x}_1^m\right)$ equals the maximizing tree assigned to the root,

$$\hat{\cal{T}}_{BIC}\left( \hat{x}_1^m\right)={\cal{T}}_{\emptyset}^d(\hat{x}_1^m).$$

\end{Propsubsub}

{\bf Proof} See Appendix.

        \begin{Obssubsub}  Once we have the estimate  $\hat{\lambda}$, associated with the tree $\cal{T}$, it is straightforward to obtain a Viterbi Algorithm  version, adapted to TSCM and TPCM, to estimate an optimal sequence of the hidden process $\boldsymbol{X}$.
\end{Obssubsub}

\section{Simulations and sensitivity analysis concerning the random noise}

In this section, we present some simulations to evaluate the methodology proposed in this work. In these simulations, we are interested in evaluating the impact on the estimation of the parameters of the hidden stochastic process $\boldsymbol{X}$, $\boldsymbol{\lambda}$, as we increase the degree of contamination $\epsilon$, for both models TSCM and TPCM. We consider simulations with sample sizes $T= 10.000$ and $30.000$ with $100$  Monte Carlo replications. The contamination  parameter  $\epsilon$ ranged from $ 0.01$ to $0.99$ with steps of $0.01$. In order to allow such a refinement in the parameter space of the random noise, we decided to use a binary alphabet to decrease the time of simulations. But we stress that there is no restriction on the methodology concerning the use of larger alphabets.
According to the proposed methodology, we used the Baum-Welch algorithm for the estimation of the parameters $\boldsymbol{\lambda}^*$ and the BIC bootstrap algorithm. 

The section is organized as follows: we present two simulation scenarios with very different trees structures of branches. For each scenario, we apply TSCM and TPCM,  in order to evaluate the parameter estimates behavior as we increase the degree of contamination of the sample.\subsection{First Scenario }

In this scenario, we chose a VLMC $\boldsymbol{X}$ of order $k=3$  with  context tree $\cal{T}$ showed in Figure \ref{figsimula} and probability transition matrix given in Table \ref{Tabsim}. We chose very different values ​​for the transition probabilities, ranging from 0.05 to 0.87, in order to observe that the behavior of the estimates does not change depending on the value chosen.

\begin{figure}[!ht]
$$\xymatrix{
                    &                      &                       &    \bullet\ar[ld]\ar[rd]  &                        &           &\\
                    &                      &*+[F]{0}\ar[ld]\ar[rd] &                           &      *+[F]{1}          &           &\\
                    &     *+[F]{00}    &                       &*+[F]{10}\ar[ld]\ar[rd]&                        &           &\\
                    &                      &    *+[F]{010}     &                           &    *+[F]{110}      &           &
}$$
\caption{Context tree  $\cal{T}$ of a VLMC $\boldsymbol{X}$ of order $k=3$.}
\label{figsimula}
\end{figure}

\begin{table}[!ht]
\caption{Transition probability matrix of the VLMC $\boldsymbol{X}$.}
\begin{center}
\begin{tabular}{l|l|l}
\hline
\multicolumn{1}{c|}{$\omega$} & \multicolumn{1}{c|}{$P(0|\omega)$} & \multicolumn{1}{c}{$P(1|\omega)$} \\ 
\hline
\multicolumn{1}{c|}{010} & \multicolumn{1}{c|}{0.05} & \multicolumn{1}{c}{0.95} \\ 
\multicolumn{1}{c|}{110} & \multicolumn{1}{c|}{0.87} & \multicolumn{1}{c}{0.13} \\ 
\multicolumn{1}{c|}{00} & \multicolumn{1}{c|}{0.27} & \multicolumn{1}{c}{0.73} \\ 
\multicolumn{1}{c|}{1} & \multicolumn{1}{c|}{0.38} & \multicolumn{1}{c}{0.62} \\ 
\hline
\end{tabular}
\end{center}
\label{Tabsim}
\end{table}

For sample sizes 10000 and 30000, we obtain accurate estimates of the parameters. We also notice that, as the contamination noise increases, the estimate of the noise parameter becomes closer to the true value, even for small samples. On the other hand, the variability of the estimates decreases as the sample size increases, as expected, and the estimates become more precise. For the TPCM model, we observe that the transition probabilities of the observed process are increasingly close to zero as the noise parameter increases, even for large samples,  which makes the estimation of the noise parameter difficult.

\footnotesize{
\begin{table}[!ht]
\caption{Estimates of some noise parameter for TSCM and TPCM.}
\begin{center}
\begin{tabular}{l|ll|ll}
\hline
\multicolumn{3}{c|}{N=10.000} & \multicolumn{2}{c}{N=30.000} \\ 
\hline
\multicolumn{1}{c|}{Noise} & \multicolumn{2}{c|}{Estimate} & \multicolumn{2}{c}{Estimate} \\ 
\hline
\multicolumn{1}{c|}{Real} & \multicolumn{1}{c|}{TSCM} & \multicolumn{1}{c|}{TPCM} & \multicolumn{1}{c|}{TSCM} & \multicolumn{1}{c}{TPCM} \\ 
\hline
\multicolumn{1}{c|}{0.01} & \multicolumn{1}{c}{0.019$\pm$ 0.011} & \multicolumn{1}{c|}{0.020$\pm$ 0.013} & \multicolumn{1}{c}{0.015$\pm$ 0.008} & \multicolumn{1}{c}{0.017$\pm$ 0.009} \\ 
\multicolumn{1}{c|}{0.05} & \multicolumn{1}{c}{0.055$\pm$ 0.012} & \multicolumn{1}{c|}{0.058$\pm$ 0.013} & \multicolumn{1}{c}{0.046$\pm$ 0.008} & \multicolumn{1}{c}{0.054$\pm$ 0.009} \\ 
\multicolumn{1}{c|}{0.25} & \multicolumn{1}{c}{0.256$\pm$ 0.013} & \multicolumn{1}{c|}{0.253$\pm$ 0.012} & \multicolumn{1}{c}{0.245$\pm$ 0.007} & \multicolumn{1}{c}{0.246$\pm$ 0.008} \\ 
\multicolumn{1}{c|}{0.45} & \multicolumn{1}{c}{0.457$\pm$ 0.012} & \multicolumn{1}{c|}{0.443$\pm$ 0.011} & \multicolumn{1}{c}{0.454$\pm$ 0.008} & \multicolumn{1}{c}{0.455$\pm$ 0.009} \\ 
\multicolumn{1}{c|}{0.55} & \multicolumn{1}{c}{0.557$\pm$ 0.011} & \multicolumn{1}{c|}{0.544$\pm$ 0.012} & \multicolumn{1}{c}{0.553$\pm$ 0.006} & \multicolumn{1}{c}{0.556$\pm$ 0.007} \\ 
\multicolumn{1}{c|}{0.75} & \multicolumn{1}{c}{0.742$\pm$ 0.013} & \multicolumn{1}{c|}{0.758$\pm$ 0.014} & \multicolumn{1}{c}{0.753$\pm$ 0.006} & \multicolumn{1}{c}{0.746$\pm$ 0.007} \\ 
\multicolumn{1}{c|}{0.95} & \multicolumn{1}{c}{0.954$\pm$ 0.012} & \multicolumn{1}{c|}{-} & \multicolumn{1}{c}{0.947$\pm$ 0.007} & \multicolumn{1}{c}{-} \\ 
\multicolumn{1}{c|}{0.99} & \multicolumn{1}{c}{0.986$\pm$ 0.011} & \multicolumn{1}{c|}{-} & \multicolumn{1}{c}{0.992$\pm$ 0.006} & \multicolumn{1}{c}{-} \\ 
\hline
\end{tabular}
\end{center}
\label{Tabeps}
\end{table}
 }\normalsize
\vskip -1cm	

\footnotesize{
\begin{table}[!htbp]\label{eps1}
\caption{Estimate of the transition matrix with TSCM regime for $\epsilon=0.01$}
\begin{center}
\begin{tabular}{l|l|l|l|l}
\hline
\multicolumn{3}{c|}{N=10.000} & \multicolumn{2}{c}{N=30.000} \\ 
\hline
\multicolumn{1}{c|}{$\omega$} & \multicolumn{1}{c|}{$P(0|\omega)$} & \multicolumn{1}{c|}{$P(1|\omega)$} & \multicolumn{1}{c|}{$P(0|\omega)$} & \multicolumn{1}{c}{$P(1|\omega)$} \\ 
\hline
\multicolumn{1}{c|}{010} & \multicolumn{1}{c|}{0.060$\pm$ 0.016} & \multicolumn{1}{c|}{0.940$\pm$ 0.016} & \multicolumn{1}{c|}{0.046$\pm$ 0.010} & \multicolumn{1}{c}{0.954$\pm$ 0.010} \\ 
\multicolumn{1}{c|}{110} & \multicolumn{1}{c|}{0.880$\pm$ 0.018} & \multicolumn{1}{c|}{0.120$\pm$ 0.018} & \multicolumn{1}{c|}{0.874$\pm$ 0.009} & \multicolumn{1}{c}{0.126$\pm$ 0.009} \\ 
\multicolumn{1}{c|}{00} & \multicolumn{1}{c|}{0.261$\pm$ 0.019} & \multicolumn{1}{c|}{0.739$\pm$ 0.019} & \multicolumn{1}{c|}{0.274$\pm$ 0.009} & \multicolumn{1}{c}{0.726$\pm$ 0.009} \\ 
\multicolumn{1}{c|}{1} & \multicolumn{1}{c|}{0.369$\pm$ 0.018} & \multicolumn{1}{c|}{0.631$\pm$ 0.018} & \multicolumn{1}{c|}{0.374$\pm$ 0.011} & \multicolumn{1}{c}{0.626$\pm$ 0.011} \\ 
\hline
\end{tabular}
\end{center}
\end{table}
}\normalsize

\bigskip

Table  \ref{eps1}  shows  the estimates of transition probabilities  for  a very small noise, $ \epsilon =  0.01$, and sample sizes $T=10000$ and $T=30000$. In both cases, the estimates are close to the true values. This was also expected since there was little change in the symbols of the original VLMC since $ \epsilon =  0.01$. And also note that as the sample size increases, the estimates becomes closer to the true one and variability decreases.
\begin{figure}[!ht]
\begin{center}
\includegraphics[scale=.15]{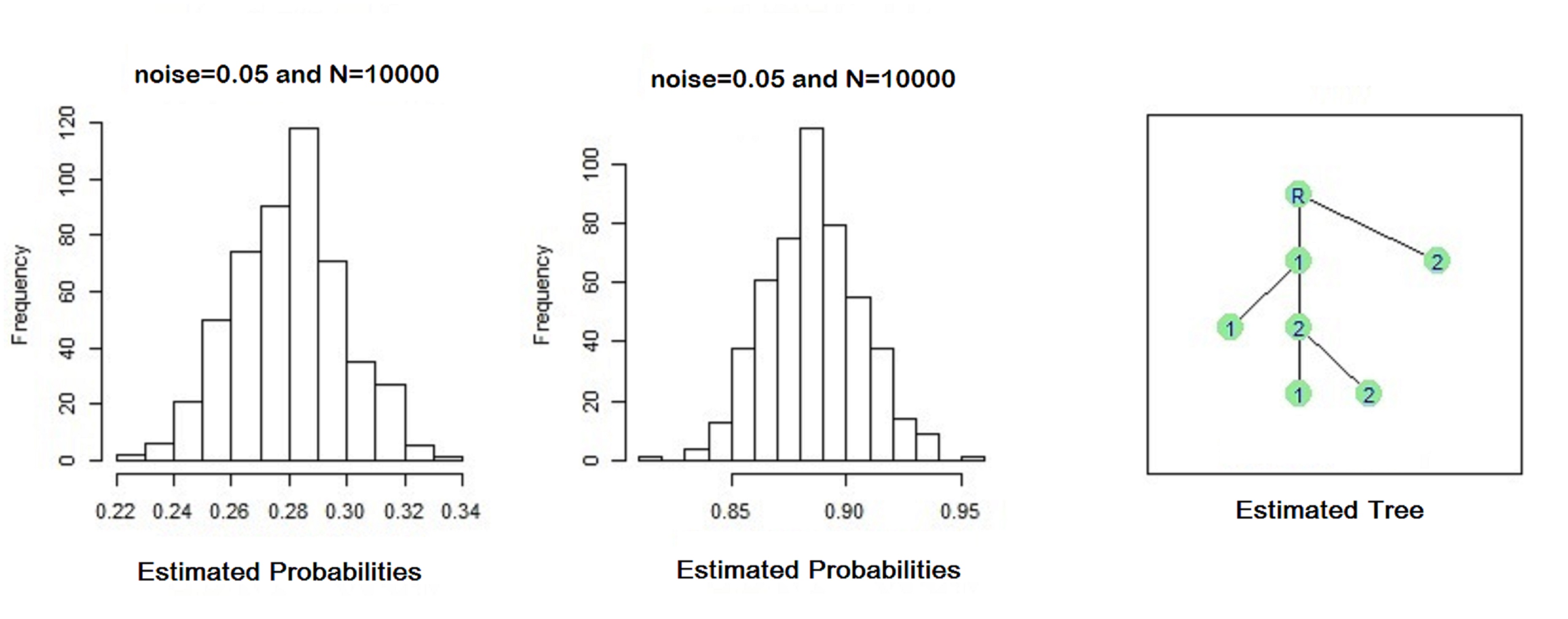}
\end{center}
\caption{Histogram of Estimated Transition Probabilities and Estimated Tree.}
\label{Figarvore}
\end{figure}

Figure \ref{Figarvore} shows evidence of normality in the behavior of the estimates of transition probabilities as the sample size increases. We only present results for estimates of transition probabilities $ 0.27$ and $0.87$ but this behavior remains the same for all other transition probabilities estimates. More than that, we notice that our methodology was able to recover the true tree. 

\footnotesize{
\begin{table}[!ht]\label{eps2}
\caption{Estimate of the Transition Matrix with TSCM regime for  $\epsilon=0.05$}
\begin{center}
\begin{tabular}{l|l|l|l|l}
\hline
\multicolumn{3}{c|}{N=10.000} & \multicolumn{2}{c}{N=30.000} \\ 
\hline
\multicolumn{1}{c|}{$\omega$} & \multicolumn{1}{c|}{$P(0|\omega)$} & \multicolumn{1}{c|}{$P(1|\omega)$} & \multicolumn{1}{c|}{$P(0|\omega)$} & \multicolumn{1}{c}{$P(1|\omega)$} \\ 
\hline
\multicolumn{1}{c|}{010} & \multicolumn{1}{c|}{0.076$\pm$ 0.020} & \multicolumn{1}{c|}{0.924$\pm$ 0.020} & \multicolumn{1}{c|}{0.068$\pm$ 0.015} & \multicolumn{1}{c}{0.932$\pm$ 0.015} \\ 
\multicolumn{1}{c|}{110} & \multicolumn{1}{c|}{0.885$\pm$ 0.021} & \multicolumn{1}{c|}{0.115$\pm$ 0.021} & \multicolumn{1}{c|}{0.862$\pm$ 0.014} & \multicolumn{1}{c}{0.138$\pm$ 0.014} \\ 
\multicolumn{1}{c|}{00} & \multicolumn{1}{c|}{0.279$\pm$ 0.022} & \multicolumn{1}{c|}{0.731$\pm$ 0.022} & \multicolumn{1}{c|}{0.275$\pm$ 0.013} & \multicolumn{1}{c}{0.725$\pm$ 0.013} \\ 
\multicolumn{1}{c|}{1} & \multicolumn{1}{c|}{0.350$\pm$ 0.021} & \multicolumn{1}{c|}{0.650$\pm$ 0.021} & \multicolumn{1}{c|}{0.362$\pm$ 0.013} & \multicolumn{1}{c}{0.638$\pm$ 0.013} \\ 
\hline
\end{tabular}
\end{center}
\end{table}
}\normalsize

Table \ref{eps2} shows estimates of the transition probabilities of the hidden process $\boldsymbol{X}$ for TSCM. We observe that the variability decreases as the sample size increases and the estimates become more accurate.
\begin{figure}[!ht]
\begin{center}
\includegraphics[scale=.2]{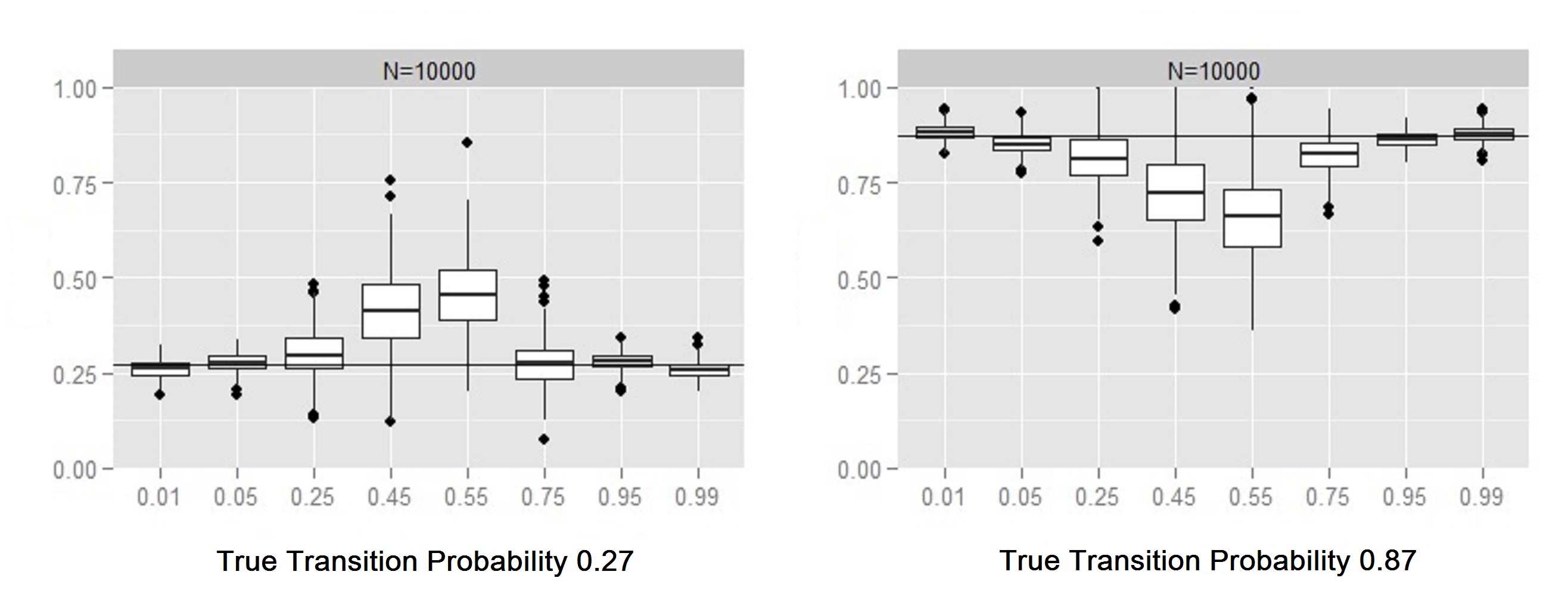}
\end{center}
\caption{Estimated transition probabilities against noise for TSCM, first scenario.}
\label{Figaruido}
\end{figure}

Figure \ref{Figaruido} shows clearly the impact of the increasing of the random noise on the estimates of transition probabilities. For noise values close to $ 0.50  $, although the noise parameters are well estimated, estimates of transition probabilities tend to be distant from the true ones and closer to $ 0.50 $. As a consequence, when the noise  is between $ 0.40  $ to $ 0.60 $, the bootstrap BIC algorithm estimates an independent model, ie, a tree with just a root. The problem occurs in the first step of the estimation procedure and not in the boostrap BIC, the Baum-Welch algorithm fails to recover the true transition probabilities in this range of random noise. It is intuitive that in TSCM with $E=\{0,1\}$ the variability of the estimators should attain higher values for noise perturbation around $0.50$  since the emission distribution is bernoulli. The high variability in this interval can lead the  Baum-Welch to fail. 

Nevertheless, if the value of the estimated noise belongs to the interval $40\%$ to  $60\%$ we can conclude that its value is well estimated, but estimates of transition probabilities are far from the true ones. But outside this range, the proposed methodology is able to provide accurate estimates for transition probabilities and also for the context tree.

\begin{figure}[h!]
\begin{center}
\includegraphics[scale=.2]{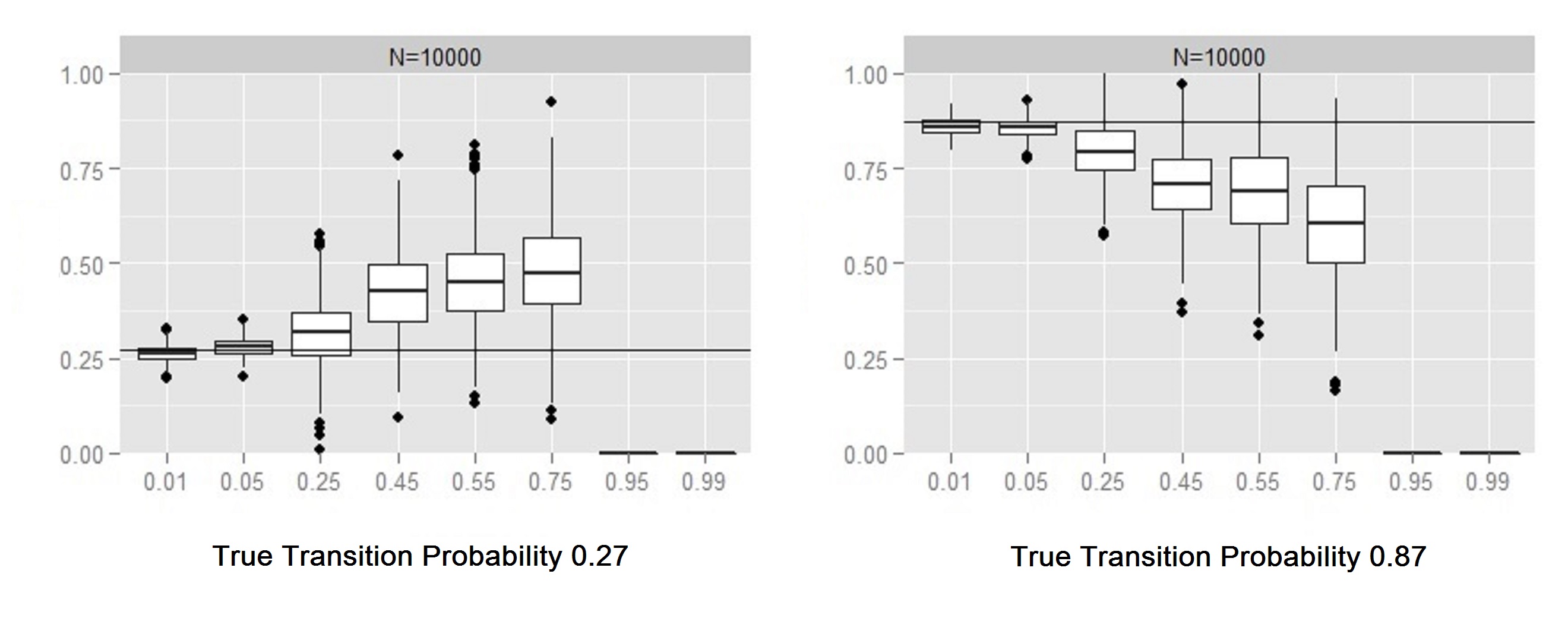}
\end{center}
\caption{Transition probabilities against  noise for TPSC first scenario.}
\label{Figaruidoprod}

\end{figure}

Table \ref{eps.prod.1} shows simulations considering TPCM regime. We observe that the estimates of the transition probabilities are very close to the true ones and become increasingly accurate by increasing the sample size, as in the TSCM. Figure \ref{Figaruidoprod} shows that, for the TPCM regime, if the contamination is smaller than $0.25$, the estimates of transition probabilities are accurate.

\footnotesize{
\begin{table}[!ht]\label{eps.prod.1}
\caption{ Estimated transition matrix  for TPCM with $\epsilon=0.01$.}
\begin{center}
\begin{tabular}{l|l|l|l|l}
\hline
\multicolumn{3}{c|}{N=10.000} & \multicolumn{2}{c}{N=30.000} \\ 
\hline
\multicolumn{1}{c|}{$\omega$} & \multicolumn{1}{c|}{$P(0|\omega)$} & \multicolumn{1}{c|}{$P(1|\omega)$} & \multicolumn{1}{c|}{$P(0|\omega)$} & \multicolumn{1}{c}{$P(1|\omega)$} \\ 
\hline
\multicolumn{1}{c|}{010} & \multicolumn{1}{c|}{0.062$\pm$ 0.015} & \multicolumn{1}{c|}{0.980$\pm$ 0.015} & \multicolumn{1}{c|}{0.055$\pm$ 0.010} & \multicolumn{1}{c}{0.945$\pm$ 0.010} \\ 
\multicolumn{1}{c|}{110} & \multicolumn{1}{c|}{0.882$\pm$ 0.018} & \multicolumn{1}{c|}{0.128$\pm$ 0.018} & \multicolumn{1}{c|}{0.871$\pm$ 0.008} & \multicolumn{1}{c}{0.129$\pm$ 0.008} \\ 
\multicolumn{1}{c|}{00} & \multicolumn{1}{c|}{0.264$\pm$ 0.019} & \multicolumn{1}{c|}{0.737$\pm$ 0.019} & \multicolumn{1}{c|}{0.277$\pm$ 0.008} & \multicolumn{1}{c}{0.723$\pm$ 0.008} \\ 
\multicolumn{1}{c|}{1} & \multicolumn{1}{c|}{0.371$\pm$ 0.018} & \multicolumn{1}{c|}{0.629$\pm$ 0.018} & \multicolumn{1}{c|}{0.376$\pm$ 0.011} & \multicolumn{1}{c}{0.628$\pm$ 0.011} \\ 
\hline
\end{tabular}
\end{center}
\end{table}
}\normalsize

We notice that, as the contamination increases, the estimates become distant from the true value, even for large samples. This is because of higher the noise in this model, most inflated zeros is the contaminated sample and more difficult is to obtain accurate estimates. Again, we only present estimates of transition probabilities $0.27$ and $0.87$ but results are similar for all other values of transition probabilities.

For a sample with size $T= 10.000 $ we are able to estimate noise values at most $ 75\% $ but the estimates are not very accurate. The methodology is able to recover the true tree for noise values smaller than $ 40\% $.  After this value, the bootstrap BIC algorithm estimates an independent model (only a root) because all transition probabilities  become closer to $ 50 \%$. However, for noise values less than $75\%$, the methodology was able to estimate the parameters of the model and to recover the true context tree $ \cal{T} $.

\subsection{Second  scenario }

In this scenario, we chose a tree with a larger order and a more complex tree structure, $\boldsymbol{X}$ is a renewal process. And as in the first scenario, the values of the transition probabilities are very different, ranging from $0.1$ to $0.83$.

Table \ref{Tabsimdois} shows the transition probability matrix associated with the process $\boldsymbol{X}$.
\begin{table}[!ht]\label{Tabsimdois}
\caption{ Transition probability matrix associated to $\boldsymbol{X}$}.
\begin{center}
\begin{tabular}{l|l|l}
\hline
\multicolumn{1}{c|}{$\omega$} & \multicolumn{1}{c|}{$P(0|\omega)$} & \multicolumn{1}{c}{$P(1|\omega)$} \\ 
\hline
\multicolumn{1}{c|}{0000} & \multicolumn{1}{c|}{0.10} & \multicolumn{1}{c}{0.90} \\ 
\multicolumn{1}{c|}{1000} & \multicolumn{1}{c|}{0.50} & \multicolumn{1}{c}{0.50} \\ 
\multicolumn{1}{c|}{100} & \multicolumn{1}{c|}{0.83} & \multicolumn{1}{c}{0.17} \\ 
\multicolumn{1}{c|}{10} & \multicolumn{1}{c|}{0.25} & \multicolumn{1}{c}{0.75} \\ 
\multicolumn{1}{c|}{1} & \multicolumn{1}{c|}{0.25} & \multicolumn{1}{c}{0.75} \\ 
\hline
\end{tabular}
\end{center}
\end{table}

\newpage
\noindent The context tree  $\cal{T}$ associated to $\boldsymbol{X}$ is shown in Figure \ref{figsimuladois} (order $k=4$).

\begin{figure}[!ht]
$$\xymatrix{
 &                      &                       &                           &                        &\bullet\ar[ld]\ar[rd]  &          &\\
 &                      &                       &                           &*+[F]{0}\ar[ld]\ar[rd]  &                       & *+[F]{1} &\\
 &                      &                       &*+[F]{00}\ar[ld]\ar[rd]    &                        & *+[F]{1}              &          &\\
 &                      &*+[F]{000}\ar[ld]\ar[rd]&                          &    *+[F]{1}            &                       &          &\\
 &   *+[F]{0000}        &                       & *+[F]{1}                    &                        &                       &        &\\
}
$$
\caption{Context tree $\cal{T}$ associated to $\boldsymbol{X}$, order $k=4$.}
\label{figsimuladois}

\end{figure}

Table \ref{Tabmatmod2} shows that we obtain accurate estimates of the parameters when $T=30000$.
\footnotesize{
\begin{table}[!ht]\label{Tabmatmod2}
\caption{ Estimates of the transition probabilities  with  TSCM for$\epsilon=0.01$.}
\begin{center}
\begin{tabular}{l|l|l|l|l}
\hline
\multicolumn{3}{c|}{N=10.000} & \multicolumn{2}{c}{N=30.000} \\ 
\hline
\multicolumn{1}{c|}{$\omega$} & \multicolumn{1}{c|}{$P(0|\omega)$} & \multicolumn{1}{c|}{$P(1|\omega)$} & \multicolumn{1}{c|}{$P(0|\omega)$} & \multicolumn{1}{c}{$P(1|\omega)$} \\ 
\hline
\multicolumn{1}{c|}{0000} & \multicolumn{1}{c|}{0.132$\pm$ 0.019} & \multicolumn{1}{c|}{0.868$\pm$ 0.019} & \multicolumn{1}{c|}{0.112$\pm$ 0.012} & \multicolumn{1}{c}{0.888$\pm$ 0.012} \\ 
\multicolumn{1}{c|}{1000} & \multicolumn{1}{c|}{0.532$\pm$ 0.018} & \multicolumn{1}{c|}{0.468$\pm$ 0.018} & \multicolumn{1}{c|}{0.515$\pm$ 0.011} & \multicolumn{1}{c}{0.485$\pm$ 0.011} \\ 
\multicolumn{1}{c|}{100} & \multicolumn{1}{c|}{0.838$\pm$ 0.015} & \multicolumn{1}{c|}{0.162$\pm$ 0.015} & \multicolumn{1}{c|}{0.825$\pm$ 0.009} & \multicolumn{1}{c}{0.175$\pm$ 0.009} \\ 
\multicolumn{1}{c|}{10} & \multicolumn{1}{c|}{0.258$\pm$ 0.016} & \multicolumn{1}{c|}{0.742$\pm$ 0.016} & \multicolumn{1}{c|}{0.246$\pm$ 0.011} & \multicolumn{1}{c}{0.754$\pm$ 0.011} \\ 
\multicolumn{1}{c|}{1} & \multicolumn{1}{c|}{0.243$\pm$ 0.018} & \multicolumn{1}{c|}{0.757$\pm$ 0.018} & \multicolumn{1}{c|}{0.253$\pm$ 0.011} & \multicolumn{1}{c}{0.747$\pm$ 0.011} \\ 
\hline
\end{tabular}
\end{center}
\end{table}
}\normalsize

\begin{figure}[!ht]
\begin{center}
\includegraphics[scale=.2]{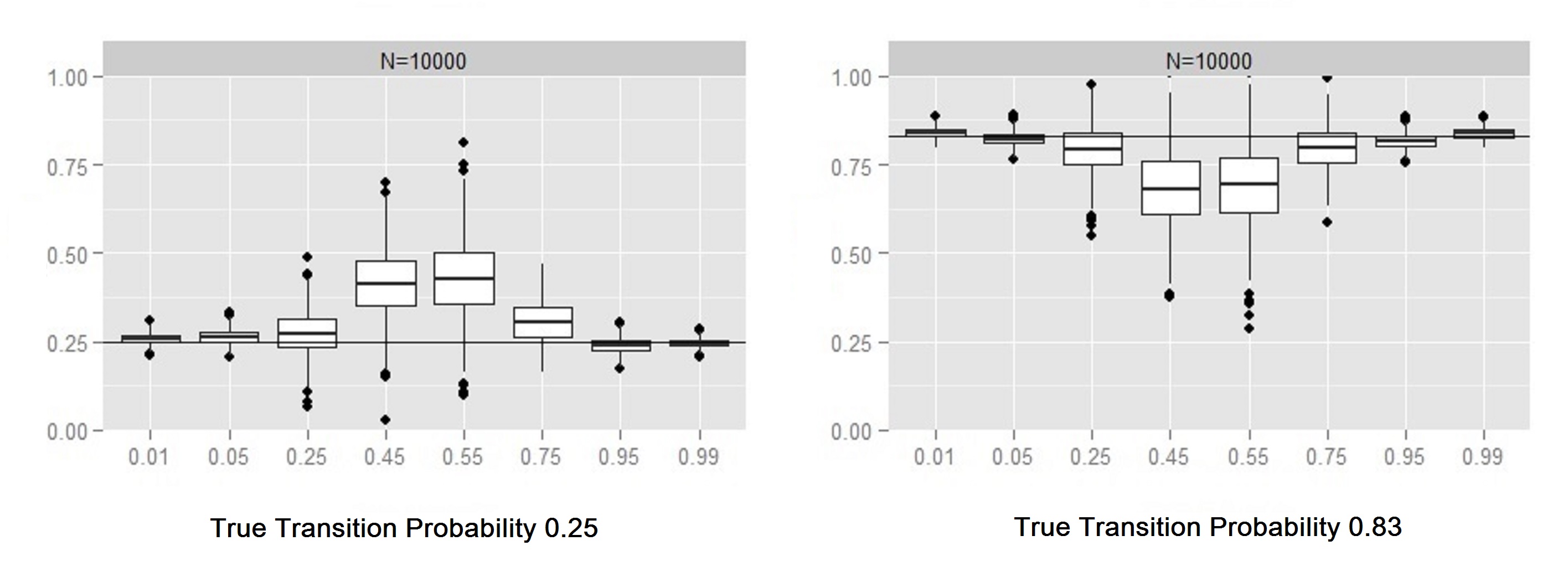}
\end{center}
\caption{Transition probabilities against random noise for TSCM, second scenario.}
\label{Figaruidomodelo2}
\end{figure}

According to the Figure \ref{Figaruidomodelo2} we observe that the estimates of the transition probabilities are accurate if the random noise is outside the interval $ 40 \% $ to $ 60 \% $, as in the first scenario. Regarding the variability of the estimates, we note that there is a range where the variability increases for a fixed sample size, but it decreases as the sample sizes increases.

\begin{figure}[!ht]
\begin{center}
\includegraphics[scale=.2]{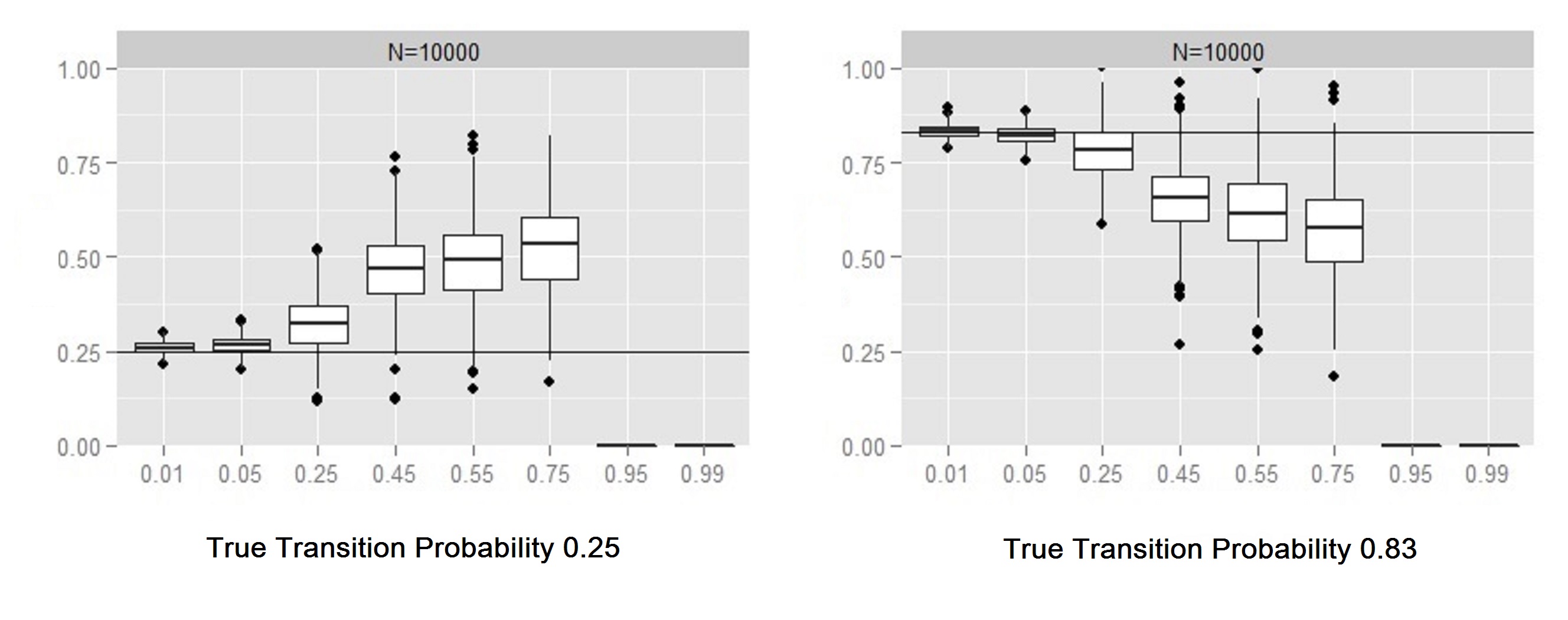}
\end{center}
\caption{Transition probabilities against random noise for TPCM, second scenario.}
\label{Figaruidomodelo3}
\end{figure}

Figure \ref{Figaruidomodelo3} shows that the estimates of the random noise and transition probabilities present the same behavior shown in the first scenario for the TPCM. 

Although the tree structure considered in scenario 2 is more complex, the results were similar to those in scenario 1.

\section{Conclusions}

In this paper, we have presented a methodology to estimate the parameters of some stochastically contaminated models. These models can be viewed as a bivariate process $(\boldsymbol{X}, \boldsymbol{Z})$ where the original, hidden, process $\boldsymbol{X}$ is as a VLMC and $\boldsymbol{Z}$ is the contaminated observed process. We considered two contamination regimes, one regime in which a random noise is added to the original value and the other contamination regime where the original value of the process is multiplied by a random noise. Our inference methodology for the parameters of these contaminated models has two steps. If the tree associated to the VLMC is finite, in the first step we rewrite  $\boldsymbol{X}$  as order $k$ Markov chain ($k_{full}$ tree) and apply the Baum-Welch EM algorithm to estimate the parameters of this transformed model. In the second step, we  proposed a bootstrap BIC in order to prune the branches of the estimated $k_{full}$ tree in the first step and, in this way, to obtain an estimate of the transition matrix of the hidden VLMC. If the tree associated to the hidden VLMC is infinite, we apply the same methodology to obtain an estimate of the parameters of the hidden tree truncated at some order $k$. 

We have shown that our bootstrap BIC  estimator for the context tree associated to $\boldsymbol{X}$ is strongly consistent under some mild conditions. We have presented simulations showing that our methodology is capable of recovering the hidden tree and the noise parameters from a contaminated sample. For samples sizes above $10.000$  the accuracy of the estimate of the noise parameter is quite satisfactory and the estimates of the transition probabilities associated to the hidden VLMC  are close to the true values,  with low variability, in a reasonable range of random noises, namely  out of $40$ to $60\%$, in the additive model, and up to  $25\%$, in the multiplicative model. Hence, if the estimate of the noise parameter is outside these ranges we can conclude that the estimates of transition probabilities associated to the hidden VLMC are reliable.

Although the simulations have been made considering an alphabet $E=\{0,1\}$, in order to decrease the time of simulations, the method can be applied to any type of emission distribution with any discrete alphabet.

\section{Appendix: {\bf Proofs }}

\noindent \textbf{ Proof of Proposition \ref{prop1}}
\begin{proof}

Let $\boldsymbol{Z}$ be a contaminated process according to a TSCM.  Without loss of generallity, we take $t=0$ and for some $\omega=a_{-l(\omega)+1}^{0}\in\cal{T}$, with $l(\omega) < T$, we have that 

\begin{eqnarray*}
P\left(Z_{0}=z_0|X_{-l(\omega)+1}^{0}=\omega\right)&=&\frac{P\left(Z_0=z_0,X_0=a_0,...,X_{-l(\omega)+1}=a_{-l(\omega)+1}\right)}{P\left(X_0=a_0,...,X_{-l(\omega)+1}=a_{-l(\omega)+1}\right)}.
\end{eqnarray*}

The event $\left\{Z_0=z_0\right\}$ can be written in terms of $\boldsymbol{X}$ and $\boldsymbol{\xi}$, according to a TSCM, as

\begin{center}
$
\left\{Z_0=z_0\right\}=\displaystyle\bigcup_{x_0,b_0=0: \atop z_0=x_0\oplus b_0}^{\left|E\right|-1}\left\{X_0=x_0,\xi_0=b_0\right\}.
$
\end{center}
Henceforth
\footnotesize{
\begin{eqnarray*}
\mathbb{P}\left(Z_{0}=z_0|X_{-l(\omega)+1}^{0}=\omega\right)&=&\frac{P\left(\displaystyle\bigcup_{{{x_0,b_0=0:}\atop {z_0=x_0\oplus b_0}}}^{\left|E\right|-1}\left\{X_0=x_0,\xi_0=b_0\right\},X_0=a_0,...,X_{-l(\omega)+1}=a_{-l(\omega)+1}\right)}{P\left(X_0=a_0,X_{-1}=a_{-1},...,X_{-l(\omega)+1}=a_{-l(\omega)+1}\right)}.
\end{eqnarray*}
}\normalsize
Note that $\left\{X_{0}=x_0,X_{0}=a_0\right\}$ are empty sets if $x_0\neq a_0$, then 

\begin{eqnarray*}
\mathbb{P}\left(Z_{0}=z_0|X_{-l(\omega)+1}^{0}=\omega\right)&=&\frac{P\left(X_0=a_{0},\xi_0=b_0,...,X_{-l(\omega)+1}=a_{-l(\omega)+1}\right)I_{\left\{z_0=a_0\oplus b_0\right\}}}{P\left(X_0=a_{0},\xi_0=b_0,...,X_{-l(\omega)+1}=a_{-l(\omega)+1}\right)}.
\end{eqnarray*}
Hence, by independence of $\boldsymbol{X}$ and $\boldsymbol{\xi}$, we have that

\begin{eqnarray}
\mathbb{P}\left(Z_{0}=z_0|X_{-l(\omega)+1}^{0}=\omega\right)&=&\frac{P\left(\xi_0=b_0\right)P\left(X_{0}=a_0,...,X_{-l(\omega)+1}=a_{-l(\omega)+1}\right)I_{\left\{z_0=a_0\oplus b_0\right\}}}{P\left(X_{0}=a_0,...,X_{-l(\omega)+1}=a_{-l(\omega)+1}\right)}\nonumber\\
&=&P\left(\xi_0=b_0\right)I_{\left\{z_0=a_0\oplus b_0\right\}}.
\end{eqnarray}

On the other hand, we have that  

\begin{eqnarray*}
P\left(Z_{0}=z_0|X_{0}=a_0\right)&=&\frac{P\left(\displaystyle\bigcup_{{{x_0,b_0=0:}\atop {z_0=x_0\oplus b_0}}}^{\left|E\right|-1}\left\{X_0=x_0,\xi_0=b_0\right\},X_{0}=a_0\right)}{P\left(X_{0}=a_0\right)}.
\end{eqnarray*}
Since the events $\left\{X_0=x_0,X_0=a_0\right\}$ are empty for all $x_0\neq a_0$, then the only remaining event is $\left\{X_0=a_0\right\}$. We notice that the events $\left\{X_0=a,\xi_0=b_0\right\}$, for each $a_0, b_0 \in E$ fixed, are mutually exclusive and $\boldsymbol{X}$ is independent of $\boldsymbol{\xi}$, then  

\begin{eqnarray}
P\left(Z_{0}=z_0|X_{0}=a_0\right)&=&\frac{P\left(X_0=a_0,\xi_0=b_0\right)I_{\left\{z_0=a_0\oplus b_0\right\}}}{P\left(X_{0}=a_0\right)}\nonumber\\
&=&P\left(\xi_0=b_0\right)I_{\left\{z_0=a_0\oplus b_0\right\}}.
\end{eqnarray}

This concludes the proof of item  \textbf{(i)}.

\bigskip

\textbf{(ii)} We want to show that  the likelihood function of the observed process $\boldsymbol{Z}$, for a sample $z_1^T$, is:
 \begin{equation*}
\mathbb{P}\left(Z_{1}^{T}=z_1^T\right)=
\displaystyle\sum_{a_{t},b_{t}\in E: \atop{1\leq t \leq T}}\displaystyle\prod_{t=1}^{T}\left[\mathbb{P}(\xi_t=b_{t})\right]\left[\mathbb{P}\left(X_{1}^{k}=a_{1}^{k}\right)\displaystyle\prod_{t=k+1}^{T}\mathbb{P}\left(X_{t}=a_{t}|X_{t-l(\omega)}^{t-1}=a_{t-l(\omega)}^{t-1}\right)\right]\displaystyle\prod_{t=1}^{T}I_{\left\{z_{t}=a_{t}\oplus b_{t}\right\}}.
\end{equation*}

Like in item {\bf (i)} in Proposition \ref{prop1} we can write the events $\left\{Z_t=z_t\right\}$ in terms of $\boldsymbol{X}$ and $\boldsymbol{\xi}$,

\begin{eqnarray*}
\mathbb{P}\left(Z_{1}^{T}=z_{1}^{T}\right)&=&\mathbb{P}\left(\displaystyle\bigcap_{1\leq t\leq T}\left[\displaystyle\bigcup_{{{a_t,b_t=0:}\atop {z_t=a_t\oplus b_t}}}^{\left|E\right|-1}\left\{X_t=a_t,\xi_t=b_t\right\}\right]\right)
\end{eqnarray*}
By the distributive property $A\cap\left\{B\cup C\right\}=\left\{A\cap B\right\}\cup\left\{A\cap C\right\}$, we have that 
\begin{eqnarray*}
\mathbb{P}\left(Z_{1}^{T}=z^{1}_{T}\right)&=&\mathbb{P}\left(\displaystyle\bigcup_{{{a_{t},b_{t}=0:}\atop {z_{t}=a_{t}\oplus b_{t}}}}^{\left|E\right|-1}\left[\bigcap_{1\leq t\leq T}\left\{X_t=a_t,\xi_t=b_t\right\}\right]\right)
\end{eqnarray*}

Since  $\left\{X_{t}=a_{t},\xi_{t}=b_{t}\right\}$ are mutually exclusive,
\begin{equation*}
\mathbb{P}\left(Z_{1}^{T}=z^{1}_{T}\right)=\displaystyle\sum_{a_t,b_t=0}^{|E|-1}\mathbb{P}\left(\displaystyle\bigcap_{{1\leq t\leq T}}\left\{X_t=a_t,\xi_{t}=b_t\right\}\right)\displaystyle\prod_{t=1}^{T}I_{\left\{z_t=a_t\oplus b_t\right\}}
\end{equation*}
 Finally, the claim follows by independence of $\boldsymbol{X}$ and $\boldsymbol{\xi}$.
\end{proof}

\noindent\textbf{Proof of Proposition \ref{propmod2}}
\begin{proof} 
Proofs of items \textbf{(i),(ii)} are analogous to the proof of proposition  \ref{prop1}, but changing the indicator function of $a\oplus b$ by $a\cdot b, \ \forall a,b\in E$.
\end{proof}

\noindent \textbf{Proof of Proposition \ref{coroCT}} 
\begin{proof} 
Let $\hat{\mathbf{A}}^*$ be a  strongly consistent estimator of the transition probability matrix of the markovian process $\boldsymbol{X}^*$, with law $\hat{P}$.
We observe that if each entry of the transition probability matrix $\hat{\mathbf{A}}^*$ , $\hat{p}(a|\omega)$, is a MLE  of the transition probability of the hidden Markov chain $\boldsymbol{X^*}$,  ${p}(a|\omega)$, $\forall a \in E, \omega \in \cal{T} $. Then for almost all realizations of the process $(\boldsymbol{X}, \boldsymbol{Z})$, we have that $\hat{p}(a|\omega)\longrightarrow p(a|\omega)$ almost surely as $m =O(T) {\rightarrow} \infty$, since the regularity conditions $A_1$ to $A_6$  in  \cite{BL} are satisfied. Hence, we only have to show that for a bootstrap sample, $\hat{x}_1^m $, of size $m=O(T)$, drawn from $\hat{P}$ fixed, for almost all realizations of the process $(\boldsymbol{X}, \boldsymbol{Z})$, the following holds
 
\begin{equation}\label{ergodic}
\frac{\hat{N}_{m}(\omega,a)}{\hat{N}_{m}(\omega)}\longrightarrow \hat{p}(a|\omega),
\end{equation}
almost surely as $m\longrightarrow\infty$.

 But since we can write 
\begin{equation}\label{NWA}
\frac{\hat{N}_m(\omega,a)}{m} = \frac{\sum_{t=k}^m 1\left\{ \hat{x}_t^{t+k}=\omega, \hat{x}_{t+k+1}=a \right\}}{m},
\end{equation}

then the random variable $\frac{\hat{N}_m(\omega,a)}{m}$, conditionally in $\hat{P}$, converges almost surely to 
\begin{center}$E(1_{\left\{ \hat{x}_t^{t+k}=\omega, \hat{x}_{t+k+1}=a\right\}}|\hat{P})= \hat{P}({\omega a})$, as $m\longrightarrow \infty$\end{center}
by the Ergodic Theorem, where $\hat{P}(\omega a)$ is the measure of the string $\omega a$ given $\hat{P}$. Analogously, we have that  
 
\begin{equation}\label{NW}
\frac{\hat{N}_{m}(\omega)}{m}\longrightarrow\hat{P}(\omega),
\end{equation}
almost surely as $m\longrightarrow\infty$.

Then,  from  \ref{NWA} and \ref{NW} the result in \ref{ergodic}  follows.

\end{proof}

\noindent {\bf Proof of the \textbf{Theorem} \ref{teoCT}}
\begin{proof}

 Lemmas 3.1 and 3.2  presented in  \cite{CT} guarantee the consistency of the BIC estimator $\hat{\cal{T}}$ in the case where the sample is obtained directly from a VLMC with tree $\cal{T}$. The only difference in our case is that we have replaced the variable $N_T(\omega,a), \omega \in \cal{T}$, $a \in E $ in \cite{CT}, which counts the frequency of the string $\omega$ followed by the symbol $a$ in the sample ${x_1^T}$, by its bootstrap version  $\hat{N}_m(\omega,a)$, but  applying Proposition \ref{coroCT} to  Lemmas 3.1 and 3.2,  with this replacement, Theorem \ref{teoCT} follows.
\end{proof}

\noindent {\bf Proof of the \textbf{Theorem} \ref{root}}
\begin{proof}
Analogously, Propositions 4.3 and Lemma 4.4 presented in  \cite{CT} guarantee the 
that $ \hat{\cal{T}}_{BIC}(x_1^n)={\cal{T}}^d_{\emptyset} $
if the sample is obtained directly from a VLMC with tree $\cal{T}$. Proposition \ref{coroCT} allows us to replace the original sample by a bootstrap sample, implying   proposition \ref{root}, by replacing $N_T(\omega,a)$, $\omega \in \cal{T}$, $ a \in  E $,  in \cite{CT} by its bootstrap version  $\hat{N}_m(\omega,a)$  in  Propositions 4.3 and Lemma 4.4 presented in  \cite{CT}.
\end{proof}

\end{document}